\documentclass[amssymb,aps,twocolumn,floats,showpacs]{revtex4-1}
\usepackage{color,graphicx,pstricks,float}
\usepackage{natbib}

\usepackage{fancybox}
\usepackage{epsfig}
\usepackage{graphicx}
\usepackage{tabularx}
\usepackage{ulem}
\begin{document}

\title{Multiple phase transitions and high-field quadrupolar order in a model for $\beta$-TeVO$_4$}

\author{Ayushi Singhania}
\author{Sanjeev Kumar}

\address{ Department of Physical Sciences,
Indian Institute of Science Education and Research (IISER) Mohali, Sector 81, S.A.S. Nagar, Manauli PO 140306, India \\
}

\begin{abstract}
Motivated by the complex behavior of the frustrated magnet $\beta$-TeVO$_4$, we study an anisotropic Heisenberg model for coupled spin-$1/2$ zigzag chains.
Using cluster mean field approach to capture quantum correlations we find, upon reducing temperature in the absence of applied field, (i) a partially ordered state, (ii) a collinear antiferromagnetic phase, and (iii) an elliptical spiral state characterized by finite vector chirality.
For finite fields, we find metamagnetic response close to saturation magnetization. We show via explicit calculations that the quadrupolar order parameter is finite in the metamagnetic regime. The exchange parameters reported in the ab-initio study of $\beta$-TeVO$_4$ are used in our study. We compare our results with those reported in recent experiments on $\beta$-TeVO$_4$ and highlight similarities as well as differences between experimental results and our cluster mean field calculations. 

\end{abstract}
\date{\today}


\maketitle

\noindent
\section{Introduction}

Interacting spin systems realized in condensed matter are well known for manifesting the surprises of quantum physics \cite{Keimer2017, Giamarchi2008}.
A microscopic understanding of quantum spin systems not only enriches us with new fundamental concepts useful across disciplines, but also opens up possibilities for next-generation technologies \cite{Balents2010, Hirschberger106, Kurumaji914}.
Quantum effects come to the fore at low temperatures in low-spin systems. Lower dimensionality and frustrating interactions further reduce the tendency of spin systems to acquire conventional magnetic order, thereby promoting exotic quantum phenomena \cite{Balents2010, Ramirez1994, Chandra1991, Hirobe2016, Lee2002, Starykh2014}. Often, the interplay among frustrating interactions, lower dimensionality and the intrinsic quantum nature of the degrees of freedom results in unexpected states and complex magnetic phase diagrams \cite{Vasiliev2018a, Lee2010}.

Weakly coupled frustrated spin-$1/2$ chains represent a class of systems possessing all of the above ingredients. These are realized in
various quasi-one-dimensional magnets, such as, linarite \cite{Linarite-PRL, Linarite-expt}, NaCuMoO$_4$(OH) \cite{expt-prb}, and LiCuVO$_4$ \cite{LiCu-expt, Enderle_2005}. 
In recent years, $\beta$-TeVO$_4$ ($\beta$TVO) has emerged as another model magnetic material for studying coupled frustrated spin chains. Low-temperature magnetic phases and phase transitions in $\beta$TVO are uncovered via a combination of thermodynamic measurements, as well as neutron scattering and NMR experiments \cite{Savina2011d, Pregelj2015e, Pregelj2016d, Weickert2016f, Pregelj2018d}.
Three magnetic transitions upon cooling, (i) paramagnet to spin density wave (SDW) at T$_{N1}=4.65$~K, (ii) SDW to spin-stripe at 
T$_{N2}=3.28$~K, and (iii) spin-stripe to an elliptical-spiral or vector chiral order at T$_{N3}=2.28$~K, are reported.
Applied field versus temperature phase diagrams were obtained via a combination of specific heat, magnetization and magnetostriction measurements \cite{Weickert2016f}. Discontinuous changes in magnetization with applied field close to saturation were reported \cite{Weickert2016f, Pregelj2015e}.
These discontinuities have been proposed as a realization of theoretically predicted spin-nematic or quadrupolar phase. Recent investigations  report that the high-field phase is magnetically ordered with nematic order possibly present in a very narrow range of applied field \cite{PRB_2019}. Experiments have reported that the intermediate spin-stripe phase hosts unusual elementary excitations called {\it wigglons} \cite{Pregelj2019} whereas at lower temperatures a coexistence of spinons and magnons is expected \cite{Pregelj2018d}.

Motivated by the rich magnetic behavior of $\beta$TVO, we study a minimal anisotropic Heisenberg model for coupled zigzag chains in two dimensions. In order to capture magnetic field and temperature dependence of various thermodynamic quantities, we make use of the cluster mean field (CMF) approach for our investigations. Using model parameters reported in ab-initio study of the material, we find: (i) a partially ordered state with zero ordered moments on alternate sites, (ii) a collinear antiferromagnet, (iii) an elliptical spiral state with finite vector chirality, and (iv) unusual metamagnetic behavior close to saturation magnetization. Some of these features, e. g., the VC ground state, the SDW order and the metamagnetic jumps in magnetization are consistent with the experimental data.

In order to place our results in a proper context, we summarize the existing theoretical work that aims to understand the frustrated spin-1/2 chain magnets in general, and $\beta$TVO in particular. Isolated zig-zag spin chains with nearest neighbor (nn) FM and next nn (nnn) AFM interactions have been studied using DMRG, exact diagonalization, effective field theories, and coupled cluster methods \cite{Bishop1998, Sirker2011, Dmitriev2008, Heidrich-Meisner2006, Vekua2007, Sato2013, Hikihara2008c, Grafe2017}. Transition from helical to SDW state driven by applied magnetic field in linarite has also been accurately described via a purely classical spin model \cite{Linarite-PRL}.
Most of these studies focus on ground state phase diagrams in the plane of interaction strength ratio and applied field. Spiral phases and nematic states close to saturation field have been reported in these theoretical investigations \cite{Sato2013, Hikihara2008c, Chubukov1991}. 
The temperature dependence in the two dimensional model has remained unexplored due to the lack of suitable methods. Classical approximation for spin operators can be invoked with the argument that thermal effects take over at finite temperatures \cite{Cinti2011}. This approach can work, with the understanding that the transition temperatures and the size of ordered moments will be overestimated, in systems that show a single phase transition. However, for a system that displays multiple transitions with changing temperature a careful treatment of quantum effects becomes most important. Since the CMF approach retains quantum correlations while allowing for thermodynamic-limit calculations of various order parameters in the mean field spirit, the method is well suited for describing temperature and magnetic field dependence within a single framework.

\noindent
\section{Model and Method} 
We begin with anisotropic Heisenberg model on coupled zigzag spin-$1/2$ chains in the presence of an external magnetic field. The model is described by the Hamiltonian,

\begin{eqnarray}
H & = &  \sum_{n, j} [ J_1 ({\bf S}_{j,n} \cdot {\bf S}_{j+1, n}) + J_2 ({\bf S}_{j,n} \cdot {\bf S}_{j+2, n} + \delta_2 S^z_{j,n} S^z_{j+2,n}) \nonumber \\
&   &  + J_{2b} ( {\bf S}_{j,n} \cdot {\bf S}_{j-1,n+1} )- h_z  S^z_{j,n} ].
\label{Ham}
\end{eqnarray}

\begin{figure}[H]
\includegraphics[width=.99\columnwidth,angle=0,clip=true]{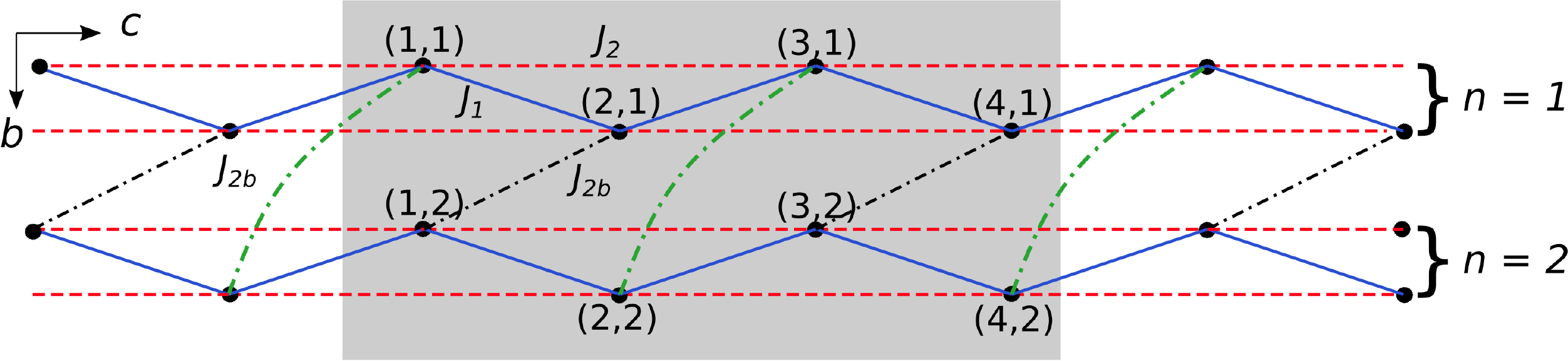}
\caption{(Color online) A schematic picture of the magnetic model for $\beta-$TeVO$_4$. The shaded area highlights the cluster within which interactions are treated exactly. The chain and site indices for each spin inside the cluster are specified. The dot-dashed lines represent connectivity due to periodic boundary condition in the $b$ direction.}
\label{cluster}
\end{figure}

\noindent
Here, ${\bf S}_{j,n}$ represents the spin-$1/2$ operator located at $n^{th}$ chain and $j^{th}$ site. $J_1$ is the nn ferromagnetic (FM) interaction, $J_2$ is the nnn antiferromagnetic (AFM) interaction and $J_{2b}$ is the AFM interchain coupling. Strength of anisotropy and applied magnetic field are denoted by $\delta_2$ and $h_z$, respectively. The parameters used in our work, following a Density Functional Theory study for $\beta-$TeVO$_4$, are: $J_1 = -26.2$K, $J_2 = 24.6$K, and $J_{2b} = 7.3$K \cite{Saul2014d, Weickert2016f}. The anisotropy value is fixed to $\delta_2 = -0.2$ throughout the study, except for results discussed in Fig. \ref{fig:S1}. We show in section III A. that exact diagonalization (ED) calculations on small clusters  miss out on most features related to conventional
magnetic order. Therefore, in this work, CMF is used as the main method to understand the nature of magnetic order and the magnetic transitions. The CMF has proven extremely useful for investigating spin models possessing magnetic frustrations of geometrical or Kitaev nature \cite{Zimmer2014, Ren2014, Zimmer2016, Gotfryd2017, Singhania2018a}.
The key idea of the method is to treat all interaction links located within the cluster exactly, and to make use of the conventional mean field decoupling, ${\bf S}_i \cdot {\bf S}_j \approx \langle {\bf S}_i \rangle \cdot {\bf S}_j + {\bf S}_i \cdot \langle {\bf S}_j \rangle - \langle {\bf S}_i \rangle \cdot \langle {\bf S}_j \rangle$, 
for interaction links connecting the cluster and the environment. 
Applying this approximation to the Hamiltonian Eq. (\ref{Ham}) leads to the cluster Hamiltonian,

\begin{eqnarray}
H_c & = &  \sum'_{n, j} [ J_1 ({\bf S}_{j,n} \cdot {\bf S}_{j+1, n}) + J_2 ({\bf S}_{j,n} \cdot {\bf S}_{j+2, n}  \nonumber \\ 
&  & + \delta_2 S^z_{j,n} S^z_{j+2,n}) + J_{2b} ({\bf S}_{j,n} \cdot {\bf S}_{j-1,n+1})- h_z  S^z_{j,n}]  \nonumber \\
&  & + \sum''_{n,j} {\bf M}^{j,n} \cdot {\bf S}_{j,n},
\label{Ham_c}
\end{eqnarray}

\noindent
where the prime over the summation sign refers to all the links contained inside the cluster, and the double-primed sum is over all those spins that have at least one interaction link outside the cluster. For our choice of cluster shown in Fig. \ref{cluster}, all eight spins contribute to the double-primed summation.
${\bf M}^{j,n}$ denotes the effective mean field vector that couples to the spin ${\bf S}_{j,n}$. The mean field vector is defined as an appropriate vector sum  $ {\bf M}^{j,n} = \sum J_{p} \langle {\bf S}_{j',n'} \rangle$, where the sum is over all spins ${\bf S}_{j',n'}$ that are located outside the cluster and are coupled to spin ${\bf S}_{j,n}$ via coupling parameter $J_{p}$, with $p = 1, 2, 2b$.
Setting up the equivalence between environment and cluster sites, in the spirit of a mean field theory, enables the closing of a self-consistency loop.
We impose periodic boundary conditions perpendicular to the chain direction ($b$-direction in Fig. \ref{cluster}), and couple the spins to mean fields along the chain ($c$-direction in Fig. \ref{cluster}). Note that the intra-cluster interaction between ${\bf S}_{3,1}$ and ${\bf S}_{2,2}$ is a consequence of periodic boundary condition along the b direction. Average of a general operator $\hat{O}$ is computed, following the standard quantum statistical approach, as,

\begin{eqnarray}
\langle \hat{O} \rangle & = & \frac{Tr~ \hat{O} ~e^{-\beta H_c}}{Tr~ e^{-\beta H_c}},
\label{av-cal}
\end{eqnarray}

\noindent
where $H_c$ is the cluster Hamiltonian defined in Eq. (\ref{Ham_c}), and the trace is over all states of the $H_c$ with converged values of the mean field parameters. Further details of the method and its extensions are available in some recent papers \cite{Zimmer2014, Ren2014, Zimmer2016, Gotfryd2017, Singhania2018a}. \\

\noindent
\section{Results and Discussions}

\subsection{Exact diagonalization}

\begin{figure}[t!]
\centering
\includegraphics[width=0.9\columnwidth,angle=0,clip=true]{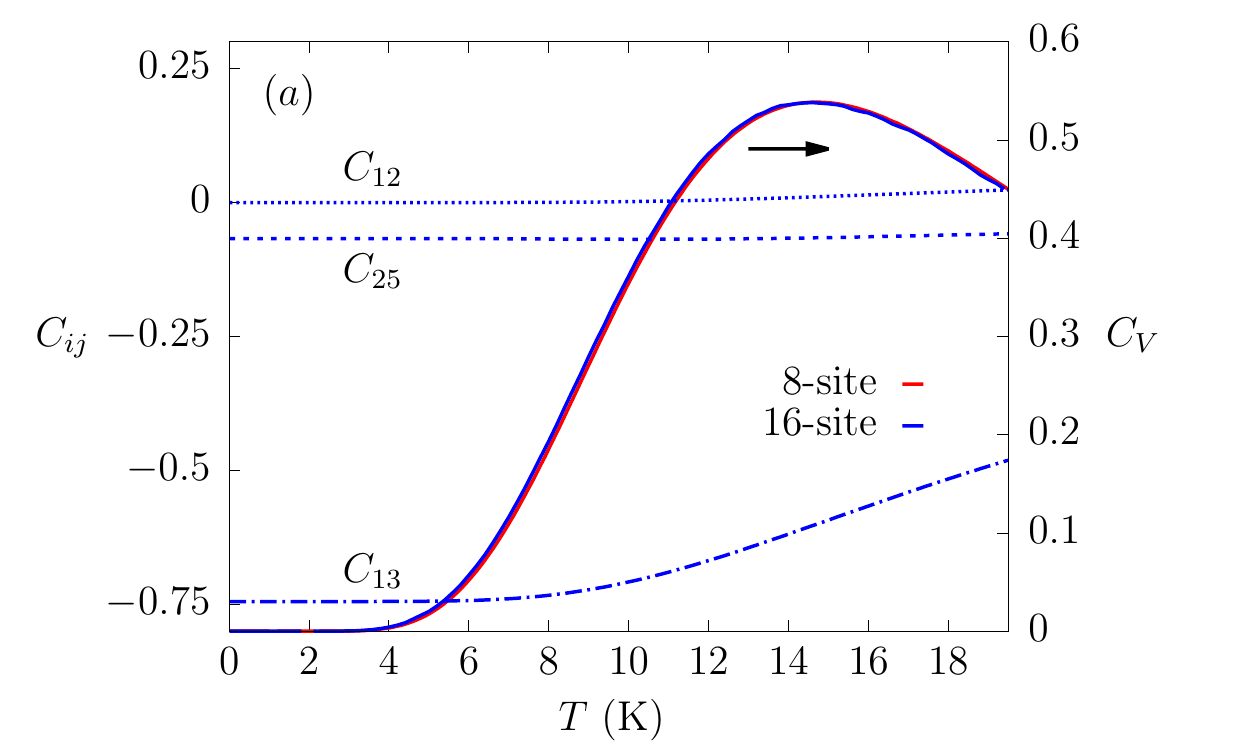} \\
\includegraphics[width=0.9\columnwidth,angle=0,clip=true]{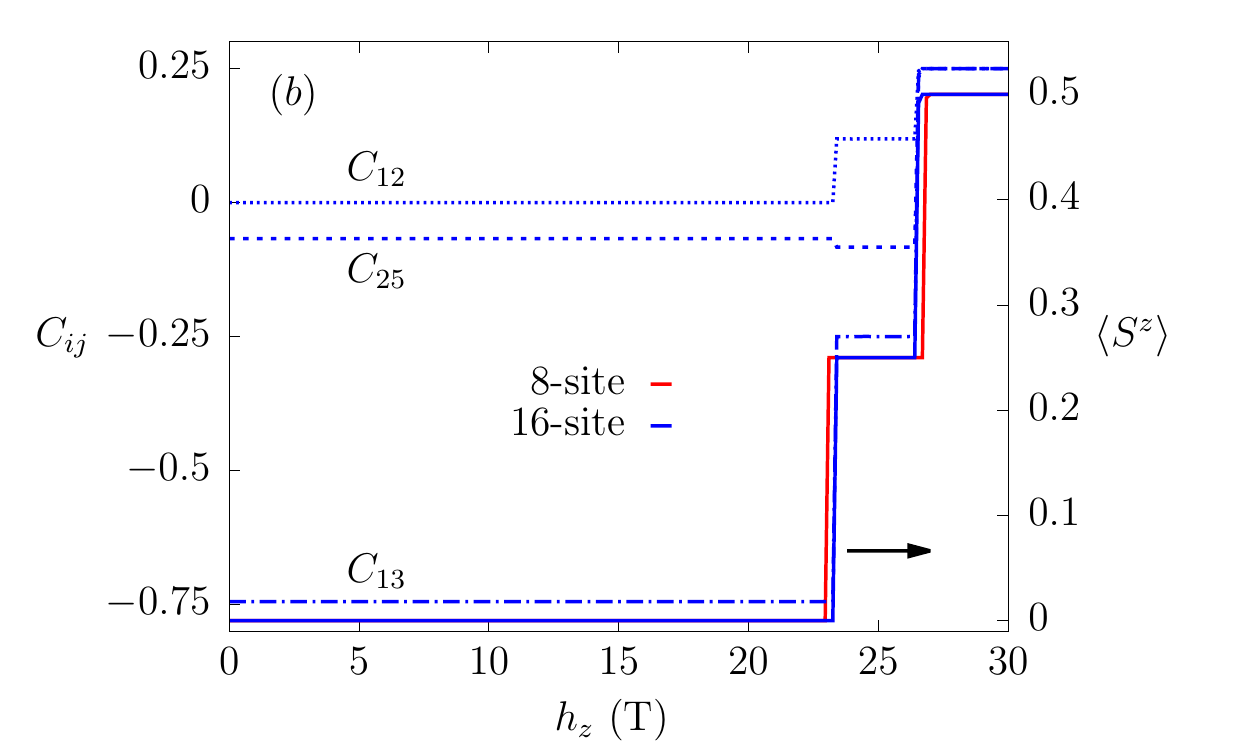}
\caption{(Color online) Solid lines display the variation of, ($a$) specific heat with temperature, and ($b$) magnetization with applied field for $8$-site and $16$-site clusters with PBC calculated using exact diagonalization. The broken lines represent variations of correlations for different spin pairs with, ($a$) temperature and ($b$) applied magnetic field for $16$-site cluster. Results in panel ($a$) are at $h_z = 0$, and those in panel ($b$) are at $T=0$. The results for spin-spin correlations for $8$-site cluster are almost identical to those for $16$-site cluster.}
\label{fig:ED}
\end{figure}	
We begin by discussing specific heat and magnetization obtained via ED calculations on finite clusters. We present results of ED for an $8$-site cluster where two zigzag chains of $4$ spins each are coupled, and a $16$-site cluster where four zigzag chains are coupled. Periodic boundary conditions (PBC) in both directions are imposed. Details of the calculations are similar to those presented in a recent study \cite{Singhania2018a}. In Fig. \ref{fig:ED} $(a)$, we show the temperature dependence of specific heat, $C_V = \frac{1}{N_c} d \langle H \rangle / dT$, where $N_c$ denotes the number of spins in the cluster. In the same panel we plot the variations of spin-spin correlations, $C_{pq} = \langle {\bf S}_{j,n} \cdot {\bf S}_{j',n'}\rangle  $ where $p = 4(n-1) + j$ and $q = 4(n'-1) + j'$ for different spin pairs. Within ED a dimer state comprising of perfect singlets on nnn sites, characterized by $C_{13} = -0.75$, emerges as the ground state. These singlet correlations gradually decrease upon increasing temperature leading to a characteristic hump in specific heat \cite{Cv_Ising, Soos_2018}. We conclude that for the Hamiltonian under consideration, ED calculations on small clusters do not support the existence of a conventional ordered magnetic state. Consequently, stand alone ED calculations are unable to provide any hint of multiple phase transitions that are observed in the experiments on $\beta$TVO. Similarly, the field dependence of magnetization displays large jumps coinciding with discontinuities in the correlation functions (see Fig. \ref{fig:ED}$(b)$). Such discontinuities associated with level crossings in the spectrum of a finite cluster will not be present in a larger system. Since ED calculations on very large systems are extremely difficult, we take an alternate approach of capturing thermodynamic limit with the help of mean field scheme. In next subsections we discuss the results obtained via CMF approach and present a comparison with available experimental data.

\subsection{Cluster Mean Field: Temperature Dependence}
\begin{figure}[t!]
\includegraphics[width=.99\columnwidth,angle=0,clip=true]{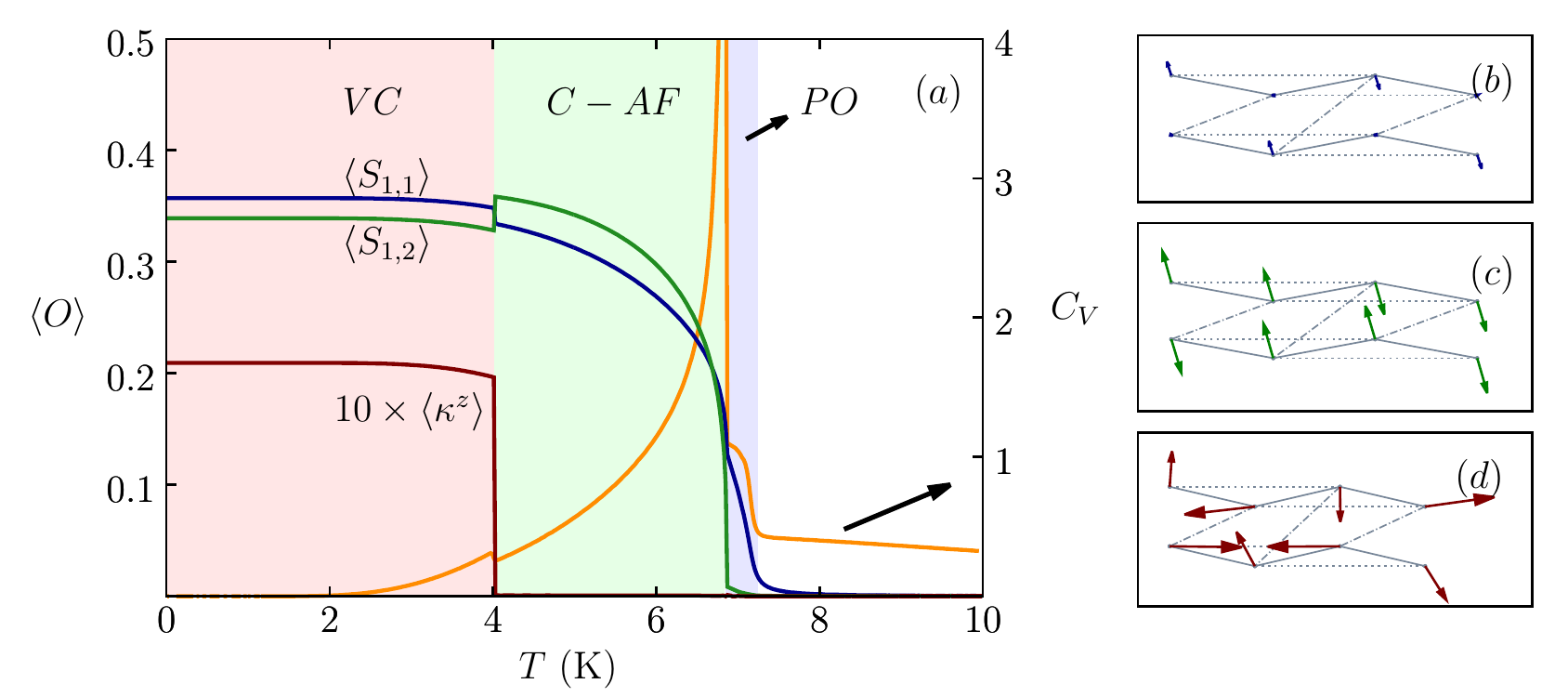}
\caption{(Color online) ($a$) Specific heat, $C_V$, vector chiral order parameter $\langle \kappa^z \rangle$, and magnitudes of self-consistent mean fields $\langle S_{1,1} \rangle$ and $\langle S_{1,2} \rangle$ as a function of temperature. The right $y$-axis is for specific heat and the left $y$-axis is for all other quantities. The vector chiral order parameter is scaled up by a factor of $10$ for clarity. Three distinct magnetically ordered states can be clearly identified with the help of these thermodynamic observables. ($b$)-($d$) Real-space patterns of mean-field spin vectors in the three phases.
}
\label{Cv}
\end{figure}

In Fig. \ref{Cv} we show the temperature dependence of specific heat $C_V$, the vector chiral (VC) order parameter $\langle \kappa^z \rangle$, and the magnitude $\langle S_{1,1} \rangle$ and
$\langle S_{1,2} \rangle$ of local spin averages at two inequivalent sites. The VC order parameter is defined as, 
\begin{eqnarray}
\langle \kappa^{z} \rangle  &=&  
\frac{1}{N_c}\sum' Im[\langle S^+_{j,n} S^-_{j+1,n} \rangle - \langle S^+_{j,n} \rangle \langle S^-_{j+1,n} \rangle ].
\end{eqnarray}
\noindent
As we will discuss below in detail, there are three distinct features in $C_V(T)$ which have an associated feature in one of the three order parameters mentioned above. 
Upon reducing $T$, one of the mean-fields, $\langle S_{1,1} \rangle$, becomes non-zero at $T = 7.2$~K while the other remains zero. This is an unusual partially ordered state where alternate sites remain quantum-disordered and are unable to develop finite magnetic moments (see Fig. \ref{Cv}($b$)). Such partially ordered states have also been reported in Kondo systems where they can be understood as arising out of the competing tendencies of Kondo screening and ordering of magnetic moments \cite{Ishizuka2013a}. PO state can be viewed as an extreme case of a SDW order where difference between magnitude of ordered local moments on neighboring sites is maximum. Indeed, the magnetic order reported in experiments on $\beta$TVO below $4.65$~K is SDW with weaker moment-size modulation and larger wavelength. The onset of the PO state is accompanied by a 'shoulder' feature in $C_V$ exactly at $T = 7.2$~K.

The second mean field parameter, $\langle S_{1,2} \rangle$, becomes finite at $T = 6.8$~K leading to a conventional magnetic order. Indeed, this is accompanied by a sharp peak in specific heat signifying a conventional long-range order (see Fig. \ref{Cv}($c$)). Upon further lowering the temperature these mean-fields approach towards their saturation values. However, a discontinuous change in both $\langle S_{1,1} \rangle$ and
$\langle S_{1,2} \rangle$ occurs at $4$~K, exactly at the temperature at which vector-chiral order parameter becomes finite confirming a change in the nature of the magnetic order. This first order phase transition manifest itself in the specific heat through a discontinuity exactly at $T = 4$~K. The ground state is therefore characterized by a finite vector chirality and unequal magnitude of magnetic moments on alternate sites (see Fig. \ref{Cv}($d$)), describing an elliptical spiral state similar to the ground state reported in experimental studies on $\beta$TVO \cite{Weickert2016f}. While the magnitude and locations of the specific-heat anomalies discussed above are likely to depend on the cluster size, the number of such anomalies remain independent of $N_c$ \cite{Singhania2018a}.

While the similarities with the experimental data \cite{Pregelj2015e, Weickert2016f} in terms of multiple phase transitions and the VC ground state become clear from the above discussion, it is important to point out the key experimental features of $\beta$TVO that are not captured in our study. We find that the ordered phase between the PO and the VC is a collinear antiferromagnet with varying moment size on alternate sites. Experiments, on the other hand, report a peculiar stripe phase with two orthogonally oriented sublattices in this regime \cite{Pregelj2016d}. Stabilizing such a phase may require inclusion of Dzyaloshinskii-Moriya interactions as well as the inter-layer coupling which we have not included in our present model Hamiltonian study.

\begin{figure}[t!]

	\includegraphics[width=.99\columnwidth,angle=0,clip=true]{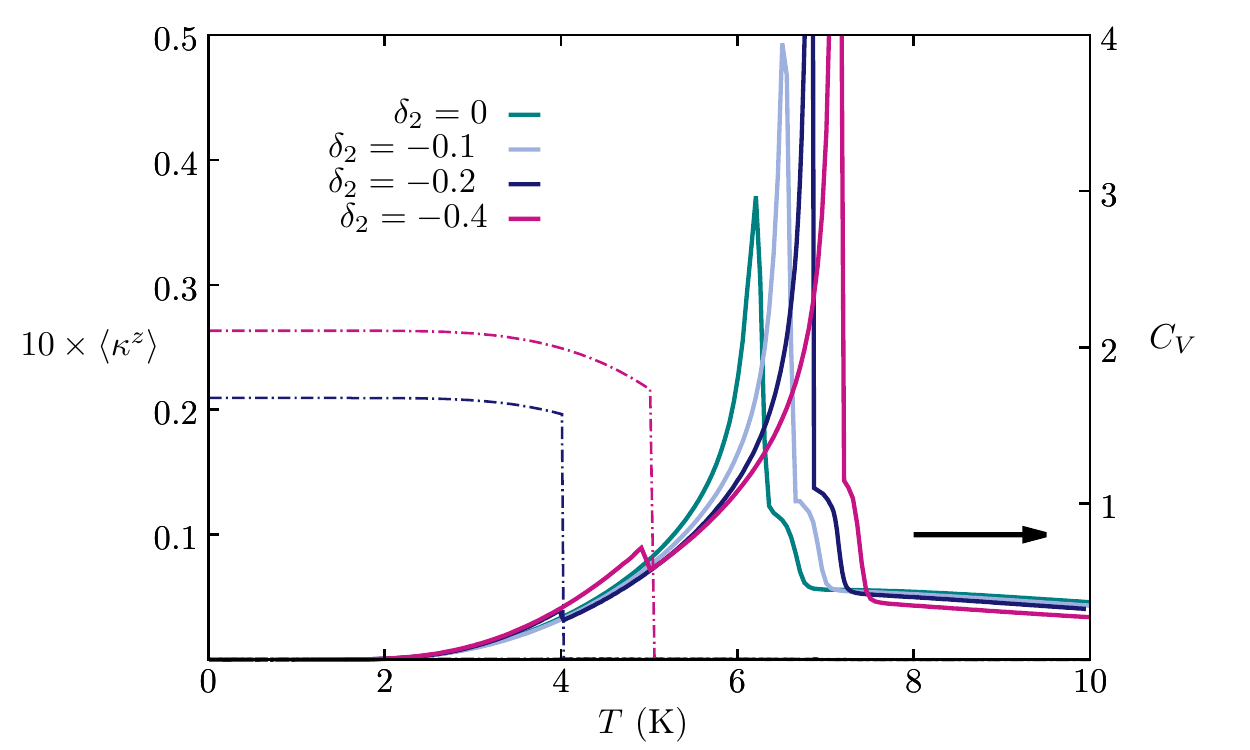}
	\caption{(Color online) Specific heat (solid lines) and vector chiral order parameter (dotted lines) as a function of temperature for different values of the anisotropy parameter $\delta_2$. Note that a significant amplitude for vector chirality appears only for $|\delta_2| \geq 0.2$.}
	\label{fig:S1}
\end{figure}

%

The appearance of VC phase as the ground state of the Hamiltonian considered in this work strongly depends on anisotropy parameter $\delta_2$. In Fig. \ref{fig:S1} we show specific heat as a function of temperature for different values of $\delta_2$. The discontinuity in the specific heat at low temperatures is absent for a fully isotropic, $\delta_2 =0$, model, and becomes prominent only for $|\delta_2| \geq 0.2$. A further increase in the strength of anisotropy parameter leads to an increase in the size of the discontinuity, as well as the value of the temperature at which the discontinuity occurs. We simultaneously track the VC order parameter. Indeed, the location of discontinuity in specific heat is correlated perfectly with the on-set of vector chirality. Therefore, we conclude that anisotropy is crucial not only to obtain the vector chiral ground state but also for the correct ordering temperature. Note that in general anisotropy in other interactions may also be present in the real material \cite{Pregelj2016d}.\\

\begin{figure}[t!]
	\centering
	\includegraphics[width=\columnwidth,angle=0,clip=true]{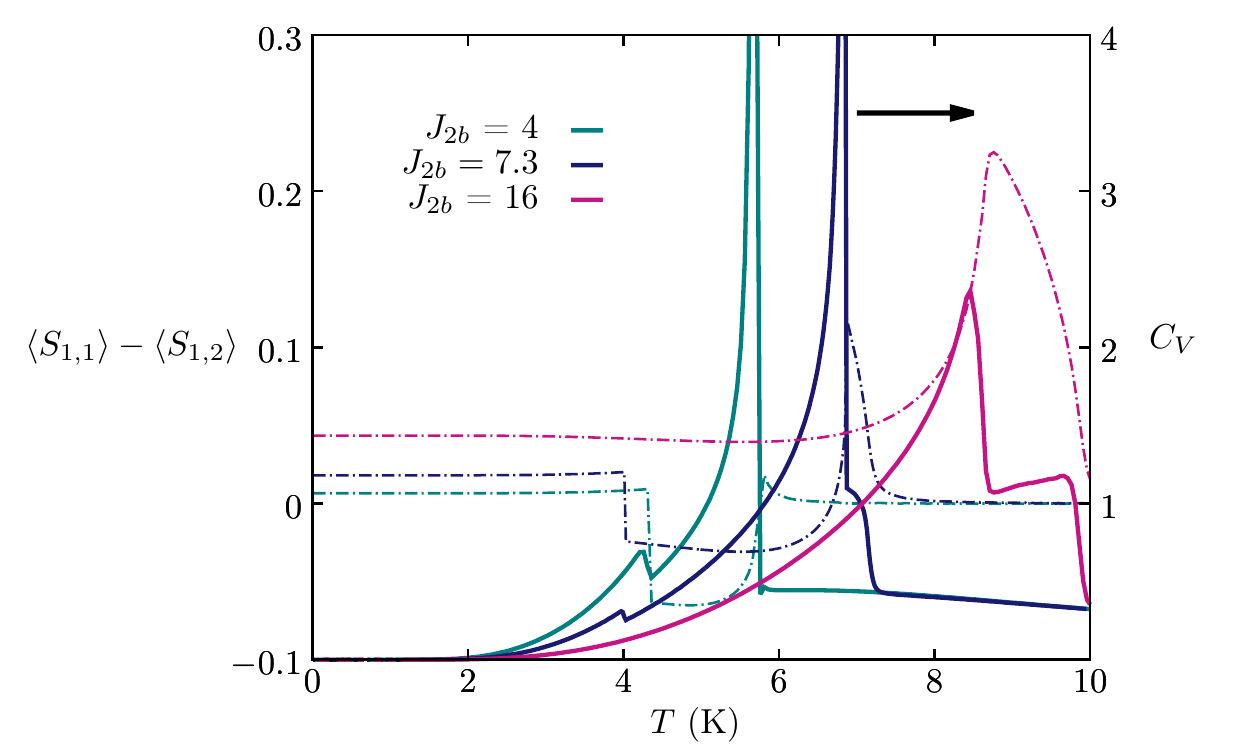}
	\caption{(Color online) Specific heat (solid lines) and the difference in the magnitude of two ordered moments (dotted lines) as a function of temperature for different values of the inter-chain coupling parameter $J_{2b}$.}
	\label{fig:S2}
\end{figure}

We also note that for the choice of parameters used for calculations, the range of stability for the partially ordered state was found to be rather narrow. To show that this state is not a result of fine tuning of parameters of the Hamiltonian, we check if the range of stability can be increased by at least one of the model parameters. 
In Fig. \ref{fig:S2}, we plot specific heat as a function of temperature for different values of the inter-chain coupling parameter $J_{2b}$. The presence of partially ordered phase is indicated by the difference between the size of local moments on two inequivalent sites. Therefore, we also show in Fig. \ref{fig:S2} the temperature dependence of $\langle S_{1,1} \rangle - \langle S_{1,2} \rangle$. Increasing the value of $J_{2b}$ leads to an increase in the window over which the specific heat displays the unusual shoulder feature. This is also followed by the large values of the moment difference $\langle S_{1,1} \rangle - \langle S_{1,2} \rangle$. However, for large values of inter-chain coupling, partially ordered state destabilizes the vector chiral ground state.
Therefore, only a suitable combination of $J_{2b}$ and $\delta_2$ allows for the appearance of both the vector chiral and the partially ordered phases.

\subsection{CMF: Magnetic Field Dependence}

\begin{figure}[t!]
\includegraphics[width=.92\columnwidth,angle=0,clip=true]{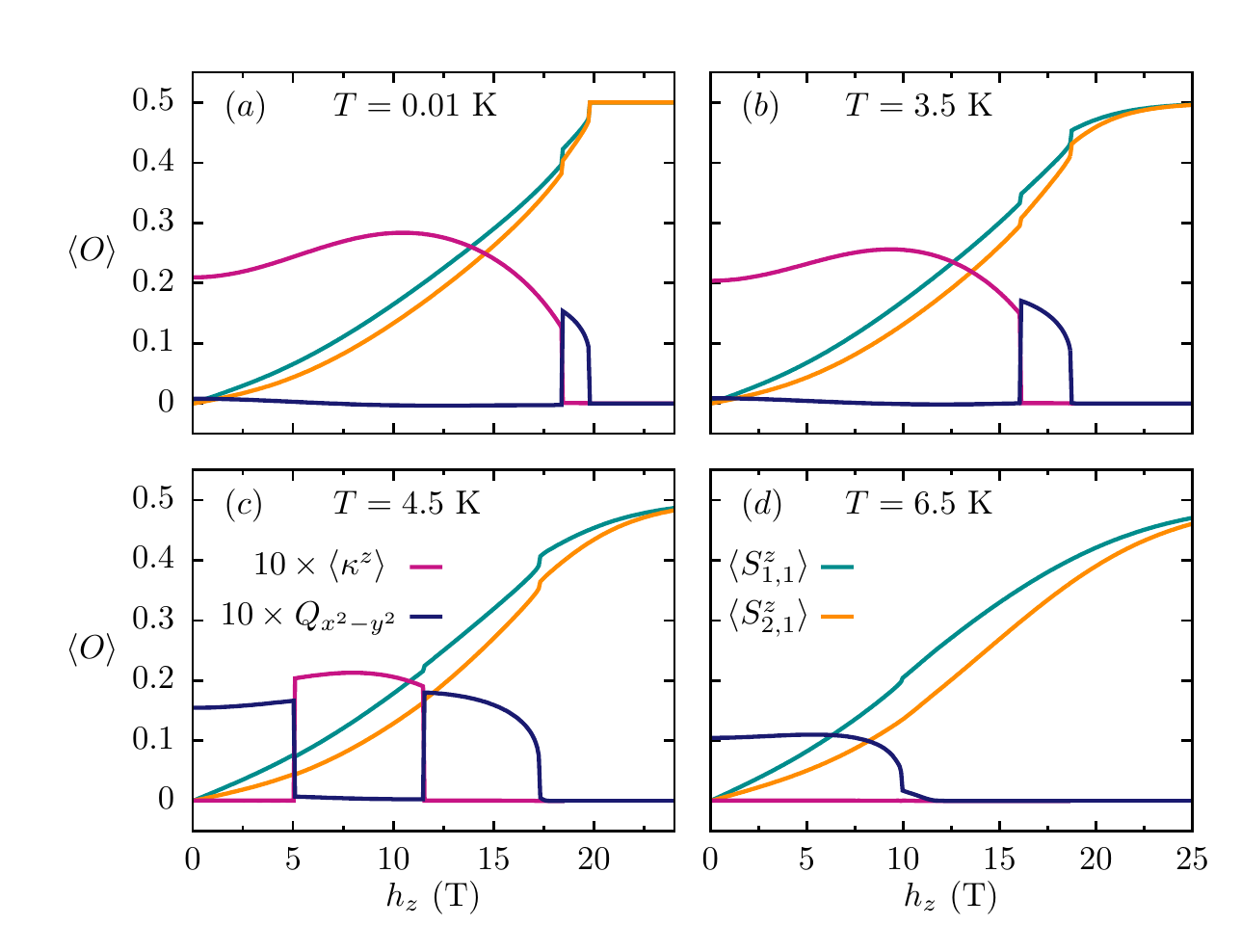}
\caption{(Color online) Component along applied field of local magnetizations, $\langle S^z_{1,1} \rangle$ and $\langle S^z_{2,1} \rangle$, on two inequivalent sites as a function of applied magnetic field for different temperatures. Magnetic field dependence of the vector chiral order parameter, $\langle \kappa^z \rangle$ and quadrupolar order parameter, $Q_{x^2-y^2}$, is also shown. The factor $k_{\rm B}/\mu_{\rm B} \approx 1.5~$ T/K is used to express $h_z$ in physical units of Tesla.}
\label{M-H}
\end{figure}

Having discussed the temperature dependence of various physical quantities in the absence of external magnetic field, we now focus on the magnetic field dependence in the low-$T$ regime. The component along external field of local magnetizations on two inequivalent sites, $\langle S^{z}_{1,1}\rangle$ and $\langle S^{z}_{2,1}\rangle$, are plotted in Fig. \ref{M-H}. Note that conversion factor of $k_{\rm B}/\mu_{\rm B}\approx1.5~$T/K is used throughout to represent the applied magnetic field values in units of Tesla.
We find a metamagnetic response in terms of discontinuities in magnetization close to saturation. We also show the field dependence of vector chiral order parameter $\langle \kappa^z \rangle$, with appropriate scale factor, in the same figure. The first jump in magnetization coincided with abrupt vanishing on VC order parameter, and  the second jump takes the system to the fully saturated FM state(Fig 6.). Therefore, we infer the existence of a new magnetic phase bounded between a fully saturated FM and the vector chiral state. These results match remarkably well with those reported in the experiments.

Recent theoretical studies based on spin-wave approach about the fully saturated FM state have suggested the presence of quadrupolar order in the metamagnetic regime.
In order to verify these predictions, we explicitly calculate quadrupolar order parameter (QOP), {\bf $Q_{x^2-y^2} = \frac{1}{N_c}\sum' Re[\langle S^+_{j,n} S^+_{j+1,n} \rangle - \langle S^+_{j,n} \rangle \langle S^+_{j+1,n} \rangle ]$}, as a function of applied field. The outcome of these calculations at different temperatures are plotted in Fig. \ref{M-H}.
Indeed, the QOP becomes finite in the metamagnetic regime (see Fig. \ref{M-H}($a$)-($b$)).
Therefore, our explicit calculations support the existing theoretical proposals regarding the presence of a quadrupolar order close to saturation magnetization \cite{Starykh2014, Zhitomirsky2010a, Sudan2009a}. Upon increasing temperature the discontinuities in magnetization become weaker in magnitude and their locations shift to lower magnetic fields and (see Fig. \ref{M-H} ($b$)-($d$)). The QOP also tracks these evolution of magnetization jumps. 
However, we also note that the QOP also becomes finite in the low-field regime for larger temperatures (see Fig. \ref{M-H}($c$)-($d$))
For $T=4.5$~K, an additional discontinuity is obtained at the on-set of the vector chiral order. This is indicative of a re-entrant behavior which becomes more clear as we discuss the complete phase diagram below. 

\begin{figure}[t!]
\includegraphics[width=.62\columnwidth,angle=0,clip=true]{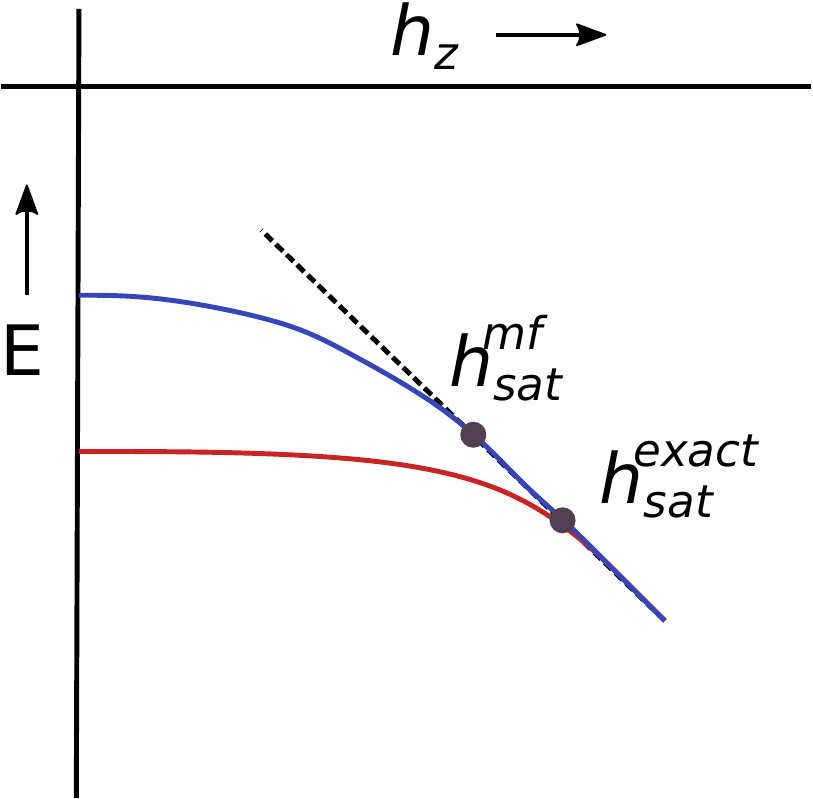}
\caption{(Color online) A schematic energy versus applied field plot demonstrating how mean field, or any approximate method, always underestimates saturation field. The lower curve is the energy obtained from exact calculations and the upper curve represents the approximate energy. The linear part is the energy of fully saturated state which is accurately captured in the mean field scheme as well. The intersection of the linear part with the exact (mean field) energy defines $h^{exact}_{sat}$ ($h^{mf}_{sat}$). 
}
\label{h-sat}
\end{figure}
 
Comparing our results with the experimental data, we note that the transition to a PM state occurs at $7.2~$K whereas the corresponding experimental value is $4.6~$K. On the other hand, the magnetic field required to obtain a fully saturated FM state, $h_z^{sat}$, is $20~$T in our study compared to $22~$T reported in experiments \cite{Weickert2016f}. It is interesting to note that while the transition temperature is significantly overestimated in the CMF approach, the saturation field is only slightly underestimated.  
We present a simple argument to justify these findings. In general, any mean field approximation suppresses thermal as well as quantum fluctuations. While both thermal and quantum fluctuations play an important role in determining the temperature at which the magnetic order is lost, only quantum fluctuations are responsible for transition to a saturated FM state at zero temperature. A more accurate estimate of $h_z^{sat}$ reflects the fact that CMF approach captures quantum fluctuations well. Furthermore, 
it can be shown that mean field approximations will always underestimate $h_z^{sat}$. The argument, presented with reference to the schematic Fig. \ref{h-sat}, is as follows: the energy of a saturated FM state is a linear decreasing function of applied field. This estimate can be accurately obtained in any approximate scheme as well. For non-saturated states, on the other hand, the ground state energy obtained via any approximate method is always higher than the exact energy. By definition, the saturation field, $h_z^{sat}$, is the value of applied field at which the energy curves for the saturated FM state (linear part in Fig. \ref{h-sat}) crosses the energy of the non saturated state. Therefore, this crossing will necessarily occur at a lower value of magnetic field for any approximate method. Note that in the schematic Fig. \ref{h-sat}, $h^{mf}_{sat} < h^{exact}_{sat}$.

\begin{figure}[t!]
\centering
\includegraphics[width=.92\columnwidth,angle=0,clip=true]{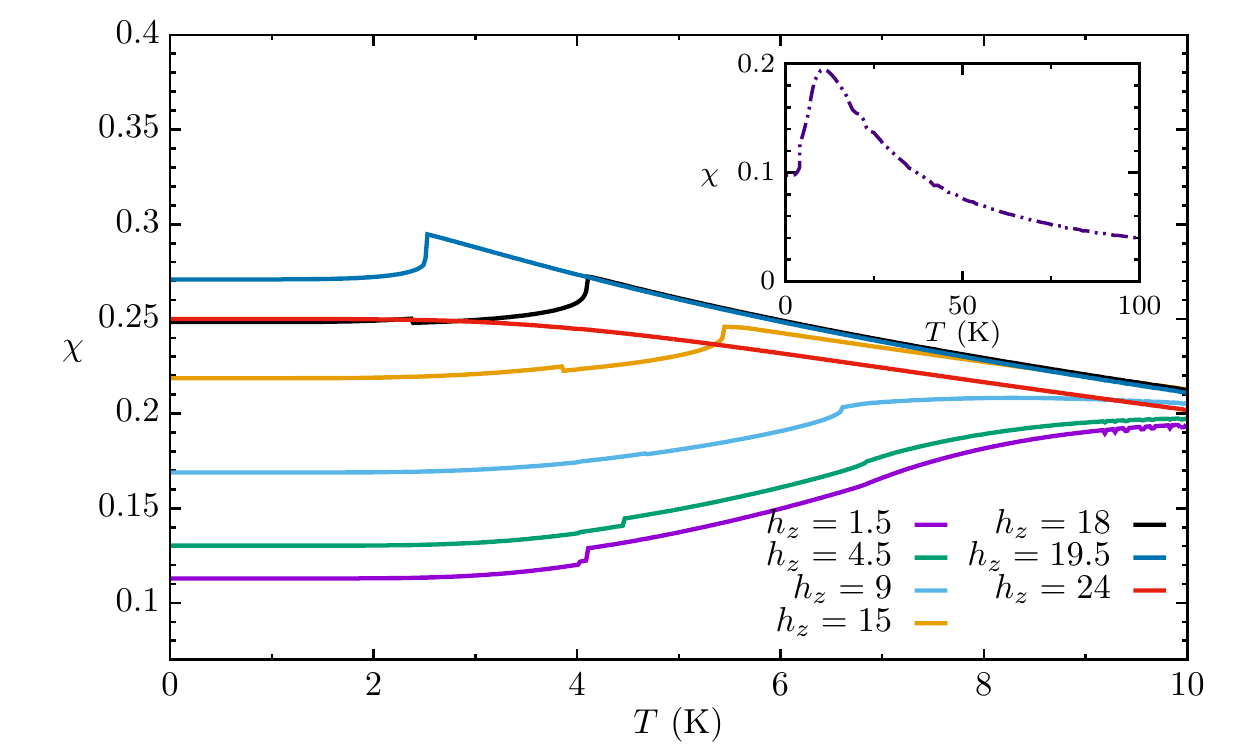}
\caption{(Color online) Magnetic susceptibility as a function of temperature for various field strengths. Inset shows susceptibility over a wider range of temperature for $h_z = 1.5 \text{T}$.}
\label{fig:S3}
\end{figure}

The magnetic susceptibility is calculated as $\chi(T)\vert_{h_z} = (dM_z / dh_z)\vert_{h_z}$ for different values of applied field $h_z$. Results are shown in Fig. \ref{fig:S3}. We find that the magnetic susceptibility shows a clear indication of the transition to the vector chiral state. However, the presence of two other transitions at higher temperatures is not very apparent looking at the magnetic susceptibility results. Interestingly, even this situation is similar to that found in experiments. The experimental susceptibility data does not provide conclusive evidence for the presence of transitions. On the other hand, the specific heat data shows clear anomalies \cite{Pregelj2015e}. For the vector chiral order, the non-monotonic evolution of the transition temperature with increasing magnetic field is correctly captured. \\

\subsection{$h_z-T$ Phase Diagram}

\begin{figure}[t!]
\includegraphics[width=.99\columnwidth,angle=0,clip=true]{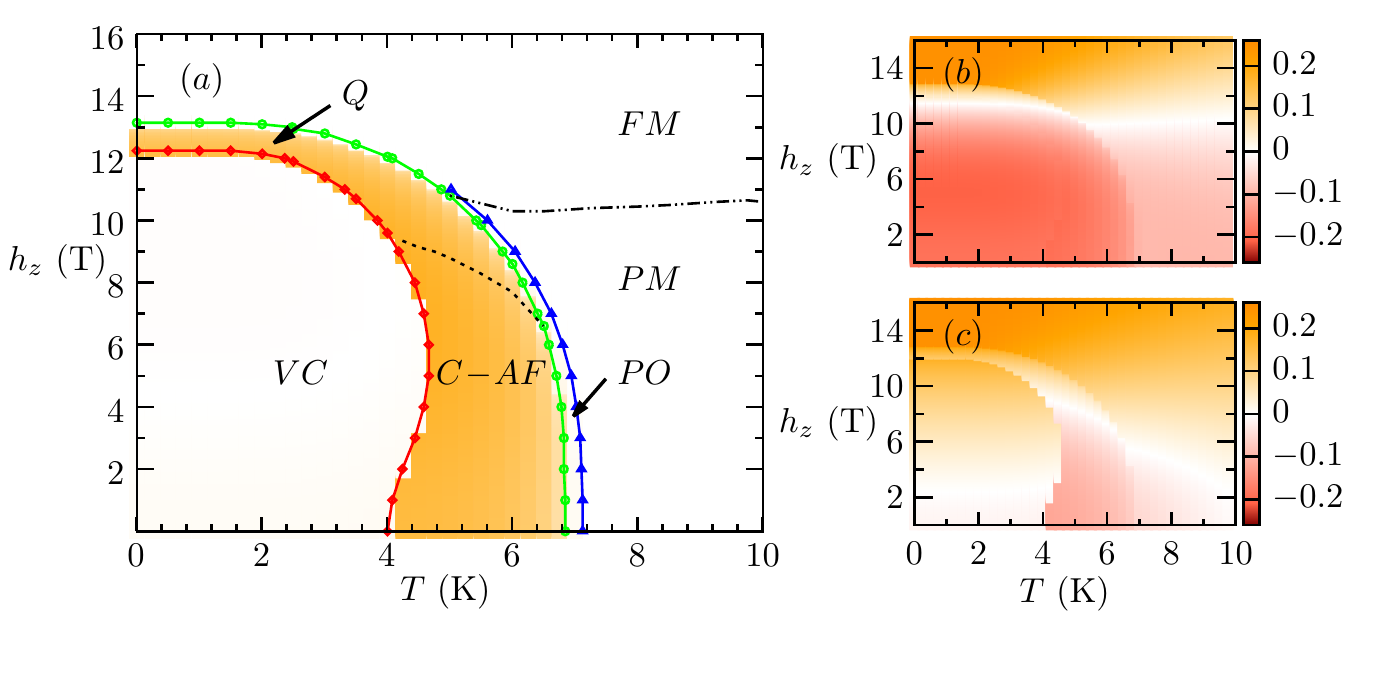}
\caption{(Color online) ($a$) The $h_z-T$ phase diagram inferred from the variations in different order parameters and the anomalies in specific heat. The color map in the background represents the value of quadrupolar order parameter $Q_{x^2-y^2}$.  ($b$)-($c$) The color map in $h_z-T$ plane of the,  ($b$) inter-chain $\langle {\bf S}_{3,1} \cdot {\bf S}_{2,2} \rangle$ ($C_{36}$), and ($c$) intra-chain $\langle {\bf S}_{2,1} \cdot {\bf S}_{3,1} \rangle$ ($C_{23}$), spin-spin correlations. Since dipolar correlations become vanishingly small in the region above the dotted line in ($a$), a pure quadrupolar order exists. 
}
\label{PD}
\end{figure}

Finally we discuss the complete $h_z-T$ phase diagram obtained in our CMF study of the anisotropic Heisenberg model. 
As discussed so far, in addition to the trivial paramagnet and fully saturated ferromagnet, we have identified four qualitatively distinct magnetic phases with varying temperature and magnetic field. These are, (i) VC, (ii) canted-AF, (iii) PO, and (iv) quadrupolar (Q) (see Fig. \ref{PD}($a$)). The boundaries between these phases are extracted from the changes in order parameters which also coincide with the anomalies in the specific heat. Intra-chain and inter-chain spin-spin correlations carry important information about the competition between different magnetic phases and provide additional insights regarding the phase boundaries. We show the color map in $h_z-T$ plane for the nn inter-chain correlation $C_{36}$ (Fig. \ref{PD}($b$)) and nn intra-chain correlation $C_{23}$ (Fig. \ref{PD}($c$)). The definition of the correlation functions as well as the site indices are as given in section III A. Qualitative changes in the inter-chain and the intra-chain correlations across different phase boundaries are clearly visible from Fig. \ref{PD}($b$) and Fig. \ref{PD}($c$).
In addition, we display the color map corresponding to the value of the QOP, $Q_{x^2-y^2}$.
The resulting phase diagram is shown in Fig. \ref{PD}. Interestingly, the region of finite $Q_{x^2-y^2}$ is bounded between two phase boundaries.
What still remains puzzling is the finiteness of QOP at low fields for which there is no support in the experimental data. We provide a resolution of this with the help of inter- and intra-chain spin-spin correlation maps.

It is interesting to note that the CMF approach allows access to both thermodynamic behavior, in the mean field spirit, as well as short range spatial correlations. This is an advantage of the CMF method over both, a fully microscopic approach such as quantum Monte Carlo, and a simple mean field approach which totally misses out on spatial correlations.
It would be interesting to check if the short range spatial correlations can provide us with additional insights regarding the behavior of the system. In particular, one may ask if the information regarding change of magnetic phases is encoded in short-range spin-spin correlations. To this end, we show in Fig. \ref{PD}($b$)-($c$) the map of correlations in the $h_z-T$ space for selected spin pairs. The intra-cluster spin-spin correlations display significant changes across various phase boundaries. The inter-chain correlations, $\langle {\bf S}_{3,1} \cdot {\bf S}_{2,2} \rangle$, become vanishingly small in the low-temperature high-field region (see Fig. \ref{PD}($b$)). Therefore, this can be seen as an effective decoupling of chains leading to destabilization of the VC order \cite{Pregelj2018d}. Furthermore, the effective reduction in dimensionality also implies a loss of long range dipolar order. Hence, this region of vanishingly small inter-chain correlations should be seen as supporting pure quadrupolar order.
Similarly, intra-chain correlation $\langle {\bf S}_{2,1} \cdot {\bf S}_{3,1} \rangle$ display a change of sign (see Fig. \ref{PD}($c$)). If the nn links in a zigzag chain get decoupled then 
the chain can be viewed as consisting of two inter-twined decoupled sublattices, and does not support long range dipolar order. 
Therefore, the region of finite QOP can be qualitatively divided into two parts with the help of the discussion above. One with coexisting long-range dipolar order and other with pure quadrupolar order. The boundary between these two regions is the dotted line in Fig. \ref{PD}($a$), which is the line of vanishing spin-spin correlations in Fig. \ref{PD}($c$).
This displays a better correspondence with the experimentally reported phase diagrams where only the high-field phase is marked as quadrupolar. Importantly, this is also consistent with the discontinuities in the magnetization which are thermodynamic signatures for the existence of an unusual, such as quadrupolar, order. It is important to note that experimental phase diagrams show appreciable dependence on the direction of applied field. Our results correspond to field applied along $b$ axis. We believe that the inter-layer couplings become essential for capturing the experimental phase diagram corresponding to the field in $a-c$ plane.

\begin{figure}[t!]
\centering
\includegraphics[width=.99\columnwidth,angle=0,clip=true]{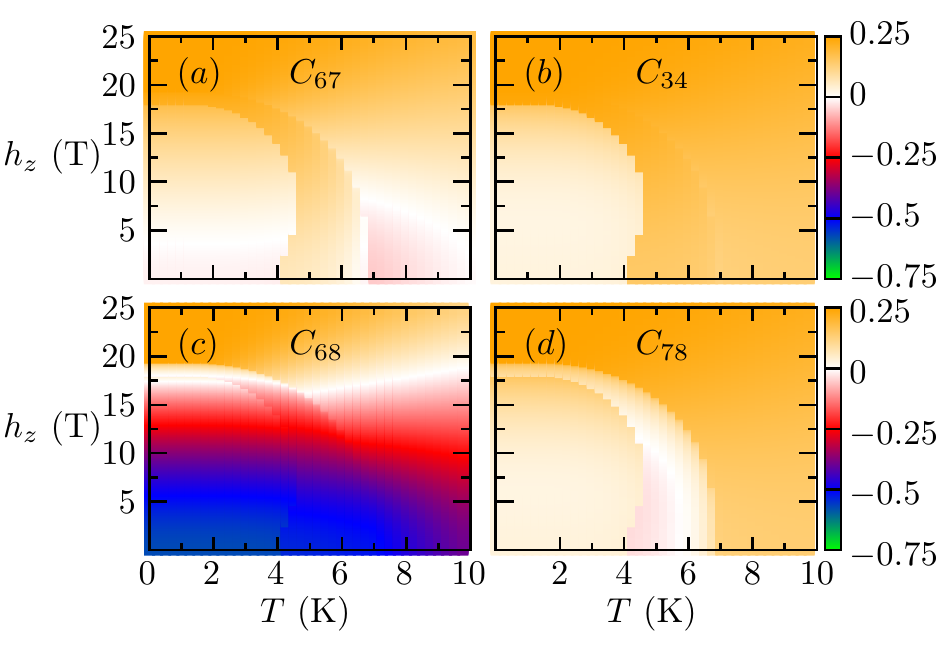}
\caption{ (Color online) ($a$)-($d$) Color map representing the value of spin-spin correlations $C{pq}$ for different spin pairs in the $h_z-T$ plane.} 
\label{fig:S7}
\end{figure}

We also display the behavior of other spin spin correlation functions in Fig. \ref{fig:S7}.
Since $|J_1| \sim J_2$ for $\beta-$ TVO, nnn spin correlations are strongest owing to their antiferromagnetic nature (see Fig. \ref{fig:S7} $(c)$). The nnn correlation $C_{68}$ gradually decreases with increasing temperature due to thermal fluctuations. Upon increasing applied field a region with vanishingly small $C_{68}$ is indicative of complete absence of dipolar order. Finally, $C_{68} = 0.25$ characterizes a field induced saturated FM state. Panels ($a$), ($b$) and ($d$) in Fig. \ref{fig:S7} display the variations of nn correlations $C_{67}$, $C_{34}$ and $C_{78}$ in that order. Given that the nn interactions are FM, these correlations have a positive value over most of the parameter regime. The re-entrant behavior as a function of applied magnetic field can also be noticed from the color maps of these nn correlation functions.

\noindent
\section{Summary and Conclusion} 
With the aim to describe complex magnetic phase diagram of $\beta$-TVO, we have investigated anisotropic Heisenberg model with nn FM and nnn AFM interactions on weakly coupled zigzag spin-$1/2$ chains. The CMF approach utilized in our study allows for an accurate treatment of short range spatial correlations, and therefore, captures the subtle competition between different possibilities of magnetic ordering. The results are obtained using realistic values of interaction parameters taken from ab-initio studies for $\beta$-TVO. We find, (i) a sequence of three phase transitions upon reducing temperature, (ii) vector chiral ground state, (iii) quadrupolar order close to saturation field accompanied by metamagnetic response, and (iv) re-entrant behavior as a function of applied field. While some of these features are consistent with the experimental data, our analysis does not capture the unusual spin-stripe state existing in the intermediate temperature regime. Additional anisotropic terms, such as Dzyaloshinskii-Moriya interaction, may be important for stabilizing the spin stripe state. Furthermore, the anisotropy of the $h_z-T$ phase diagram may crucially depend on the inter-chain couplings along the $a$ axis which is not included in the present study \cite{Weickert2016f}. We have also shown that
the relative locations of the transition temperatures can be tuned by varying the relative strengths of the coupling parameters. Nevertheless, the transition temperatures are overestimated due to the mean field nature of the method. To conclude, 
in addition to capturing certain general features of the complex magnetic phase diagram for $\beta$-TVO, our results highlight how the CMF approach can become a powerful tool in understanding the nature of magnetic order emerging at low temperatures in frustrated magnets.

\noindent
\section{Acknowledgments}
We thank Satoshi Nishimoto for useful discussions. We acknowledge the use of High-Performance Computing Facility at IISER Mohali.


%

\end{document}